\documentclass[journal]{rmaa}

\usepackage{graphicx}
\usepackage[utf8]{inputenc}

\title{The Photometric Variability of HH~30} 

\altaffiltext{1}{Centro de Radioastronomía y Astrofísica, 
    Universidad Nacional Autónoma de México.}
\altaffiltext{2}{Jet Propulsion Laboratory, Caltech.}

\author{
  Alan M. Watson\altaffilmark{1}, 
  María Carolina Durán-Rojas\altaffilmark{1},
  and
  Karl R. Stapelfeldt\altaffilmark{2}
}

\shortauthor{Watson et al.}
\shorttitle{The Photometric Variability of HH~30}

\listofauthors{
  A. M. Watson,
  M. C. Durán-Rojas,
  \& 
  K. R. Stapelfeldt
}

\fulladdresses{
\item
A.~M.~Watson and M.~C.~Durán-Rojas: Centro de Radioastronomía y
Astrofísica, Universidad Nacional Autónoma de México, Apartado Postal
3-72 (Xangari), 58089 Morelia, Michoacán, México
(a.watson@astrosmo.unam.mx, c.duran@astrosmo.unam.mx).
\item
K.~R.~Stapelfeldt: Jet Propulsion Laboratory, California Institute of
Technology, Pasadena, CA, USA (krs@exoplanet.jpl.nasa.gov).
}

\indexauthor{Watson, A. M.}
\indexauthor{Durán-Rojas, M. C.}
\indexauthor{Stapelfeldt, K. R..}

\abstract{HH~30 is an edge-on disk around a young stellar object. Previous imaging with the {\itshape Hubble Space Telescope} has show morphological variability that is possibly related to the rotation of the star or the disk. We report the results of two terrestrial observing campaigns to monitor the integrated magnitude of HH~30. We use the Lomb-Scargle periodogram to look for periodic modulation with periods between 2 days and almost 90 days in these two data sets and in a third, previously published, data set. We develop a method to deal with short-term correlations in the data. Our results indicate that none of the data sets shows evidence for significant periodic photometric modulation.}

\resumen{HH~30 es un disco visto casi de canto alrededor de un objeto estelar joven. Imágenes previas del {\itshape Hubble Space Telescope} muestran una variabilidad morfológica que posiblemente esté relacionada con la rotación de la estrella o el disco. Reportamos los resultados de dos campañas observacionales realizadas con un telescopio terrestre para monitorear la magnitud integrada de HH~30. Usamos el periodograma de Lomb-Scargle para buscar modulaciones periódicas con periodos entre 2 y casi 90 días en estos dos conjuntos de datos y en un tercer conjunto de datos previamente publicado. Desarrollamos un método para mitigar los efectos de las  correlaciones de periodo corto en los datos. Nuestros resultados indican que ninguno de los conjuntos de datos muestra evidencia de una modulación periódica en su photometría.}

\addkeyword{accretion, accretion disks}
\addkeyword{circumstellar matter}
\addkeyword{stars: individual (HH~30)}
\addkeyword{stars: pre-main sequence}

\begin{document}

\maketitle

\section{Introduction}

High-resolution images show that HH~30 is
a compact bipolar reflection nebula bisected by a dark lane (Burrows et
al.\ 1996). Its location in the L1551 molecular
cloud and similarity to the model images of Whitney \& Hartman
(1992) led immediately to the conclusion that HH~30 is an optically-thick
circumstellar disk seen almost edge-on around a young stellar object.

An interesting aspect of HH~30 is the prominent morphological
variability (Burrows et al.\ 1996;
Stapelfeldt et al.\ 1999; Cotera et al. 2001; Watson \& Stapelfeldt
2007). This variability includes changes in the contrast between the
brighter and fainter nebulae over a range of more than one magnitude;
changes in the lateral contrast between the two sides of the brighter
nebula over a range of more than one magnitude; and changes in the
lateral contrast between the two sides of the fainter nebula over a
range of about half a magnitude. It appears that the central source is
acting as a lighthouse, preferentially illuminating different parts of
the disk.

Two mechanisms have been suggested for the lighthouse. Wood \& Whitney
(1998) suggested non-axisymmetric stellar accretion hot-spots.
Stapelfeldt et al.\ (1999) suggested voids or clumps in the inner disk.
AA Tau seems to be a prototype for both mechanisms, apparently
possessing both inclined hot spots and occulting inner-disk warps, both
presumably the result of an inclined stellar magnetic dipole (Bouvier et
al.\@ 1999; Ménard et al.\@ 2003; O'Sullivan et al.\@ 2005).

The two mechanisms are likely to be periodic, as they are expected to be
tied to stellar rotation and orbital motions. Therefore, there is a hope
that we might see a periodic modulation in the integrated photometry of
HH~30, as one might expect the nebulae to be observed to be brighter
when the lighthouse beam is pointing towards the observer. In this work
we report an unsuccessful attempt to detect such a periodic modulation in
three data sets.

\section{Observations}

\subsection{Data Set 1}

We observed HH~30 with the 84 centimeter telescope of the Observatorio
Astronómico Nacional on Sierra San Pedro Mártir on 24 of the 28 nights
between 1999 January 29 and 1999 February 25. We used the SITe1 $1024
\times 1024$ CCD binned $1\times1$ with the observatory's $R$ (2 mm KG3
and 2 mm OG570) and $I$ (4 mm RG9) filters.

\begin{figure}
\centering
\includegraphics[width=\columnwidth]{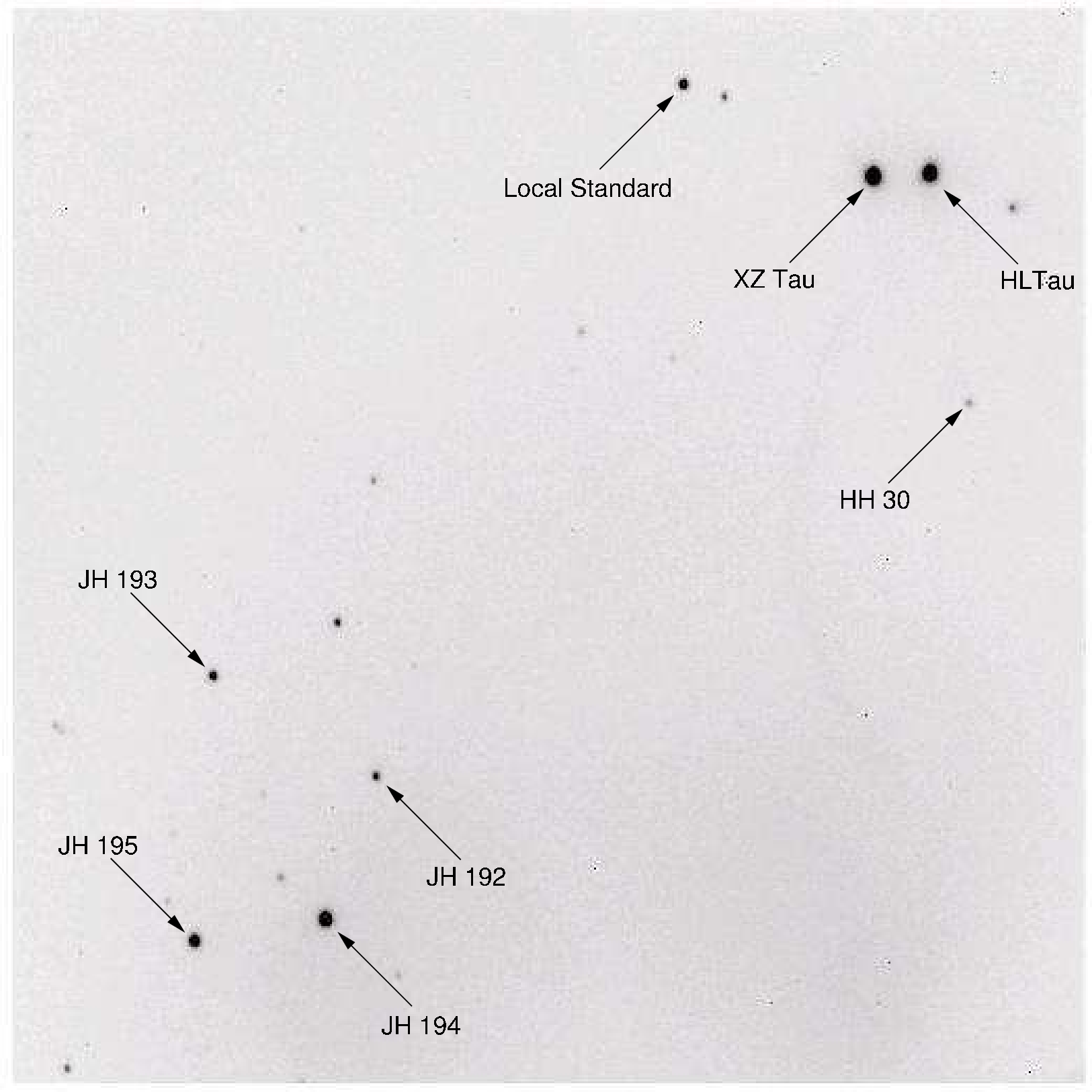}
\medskip
\caption{An image of the field of HH~30 in $I$ taken on 1999 January 29.
  North is approximately up and east is approximately left. The image is
  approximately $7'$ to a side. This pointing is typical of those used
  to obtain data set 1. The local standard is 2MASS 04314544+1814359,
  which does not vary significantly with respect to the stars JH 192,
  193, 194, and 195.}
\label{figure:field}
\end{figure}

We selected the pointing to include both HH~30 and several stars to the
south-east. A typical pointing is shown in Figure~\ref{figure:field}. We used
roughly the same pointing each night, to minimize variations in residual
flat field error. Each night we typically took two consecutive 600
second exposures in $R$ and two consecutive 300 second exposures in $I$,
although on the night of 1999 February 11 we were able to observe only in
$R$. The image quality was typically $2''$ FWHM, and we often observed
through clouds.

We reduced each image by subtracting an offset calculated from the
overscan, subtracting a residual bias image, and dividing by a twilight-sky flat field. We obtained instrumental magnitudes of all of the bright
sources in the field using aperture photometry with an object aperture
of diameter $8''$ and a sky annulus with an inner diameter of $8''$
and an outer diameter of $16''$. We averaged the instrumental magnitudes
in the consecutive images in each filter.

\begin{table*}
\centering
\setlength{\tabnotewidth}{\columnwidth}
\tablecols{8}
\caption{Stability of the Local Standard in 1999 January and February}
\label{table:local-standard}
\begin{tabular}{cccccccc}
\toprule
&&&
\multicolumn{2}{c}{$\Delta R$}&&
\multicolumn{2}{c}{$\Delta I$}\\
\cline{4-5}
\cline{7-8}
JH & 2MASS &&
$\bar m$& $\sigma_m$&&
$\bar m$& $\sigma_m$\\
\midrule
JH 192 & J04315455+1809573 && $+1.002$ & 0.013 && $+0.534$ & 0.018 \\
JH 193 & J04315915+1810391 && $+0.473$ & 0.015 && $-0.033$ & 0.019 \\
JH 194 & J04315607+1808595 && $-1.483$ & 0.009 && $-1.466$ & 0.017 \\
JH 195 & J04315981+1808517 && $-0.779$ & 0.013 && $-1.335$ & 0.016 \\
\bottomrule             
\end{tabular}
\end{table*}

We adopted the star 2MASS 04314544+1814359 as a local standard. This
star lies $131''$ north and $114''$ east of HH~30 and is marked in
Figure~\ref{figure:field}. Table~\ref{table:local-standard} shows
differential photometry of stars JH 192, 193, 194, and 195 (Jones \&
Herbig 1979 and Figure~\ref{figure:field}) against the local standard.
In Table~\ref{table:local-standard}, $\bar m$ is the mean magnitude
difference over the run and $\sigma_m$ is the empirical
estimate of the standard deviation of a single magnitude difference. The
standard deviations are all less than 2\%, and show that the local
standard did not vary significantly during the course of our
observations.

\begin{table*}
\centering\footnotesize
\setlength{\tabnotewidth}{\columnwidth}
\tablecols{9}
\caption{Data Set 1: Photometry of HH~30 from 1999 January and February}
\label{table:observations-I}
\begin{tabular}{lcccccccc}
\toprule
&&
\multicolumn{3}{c}{$\Delta R$}&&
\multicolumn{3}{c}{$\Delta I$}\\
\cline{3-5}
\cline{7-9}
\multicolumn{1}{c}{Date}&&
JD & $m$& $\sigma_m$&&
JD & $m$& $\sigma_m$\\
\midrule
1999 Jan 29 && 2451208.800 & 1.628 & 0.016 && 2451208.788 & 2.184  & 0.014 \\
1999 Jan 30 && 2451209.648 & 1.726 & 0.047 && 2451209.626 & 2.210  & 0.011 \\
1999 Jan 31 && 2451210.627 & 1.639 & 0.012 && 2451210.647 & 2.219  & 0.013 \\
1999 Feb 01 && 2451211.661 & 1.604 & 0.013 && 2451211.644 & 2.124  & 0.011 \\
1999 Feb 02 && 2451212.730 & 1.504 & 0.010 && 2451212.750 & 2.071  & 0.012 \\
1999 Feb 03 && 2451213.682 & 1.743 & 0.011 && 2451213.664 & 2.236  & 0.011 \\
1999 Feb 06 && 2451216.642 & 2.081 & 0.011 && 2451216.629 & 2.746  & 0.016 \\
1999 Feb 07 && 2451217.618 & 2.069 & 0.011 && 2451217.603 & 2.746  & 0.019 \\
1999 Feb 09 && 2451219.632 & 2.171 & 0.012 && 2451219.620 & 2.898  & 0.019 \\
1999 Feb 11 && 2451221.672 & 2.089 & 0.020 &&\nodata      &\nodata &\nodata\\
1999 Feb 12 && 2451222.675 & 2.114 & 0.010 && 2451222.659 & 2.710  & 0.015 \\
1999 Feb 13 && 2451223.756 & 1.769 & 0.008 && 2451223.740 & 2.320  & 0.011 \\
1999 Feb 14 && 2451224.686 & 1.702 & 0.007 && 2451224.662 & 2.276  & 0.010 \\
1999 Feb 15 && 2451225.617 & 1.627 & 0.008 && 2451225.605 & 2.195  & 0.010 \\
1999 Feb 16 && 2451226.686 & 1.855 & 0.010 && 2451226.662 & 2.442  & 0.013 \\
1999 Feb 17 && 2451227.628 & 1.978 & 0.012 && 2451227.619 & 2.586  & 0.017 \\
1999 Feb 18 && 2451228.633 & 1.946 & 0.012 && 2451228.615 & 2.574  & 0.016 \\
1999 Feb 19 && 2451229.688 & 1.849 & 0.010 && 2451229.664 & 2.405  & 0.012 \\
1999 Feb 20 && 2451230.681 & 1.653 & 0.008 && 2451230.665 & 2.166  & 0.010 \\
1999 Feb 21 && 2451231.686 & 1.741 & 0.009 && 2451231.669 & 2.270  & 0.010 \\
1999 Feb 22 && 2451232.667 & 1.777 & 0.034 && 2451232.653 & 2.332  & 0.015 \\
1999 Feb 23 && 2451233.698 & 1.859 & 0.043 && 2451233.685 & 2.349  & 0.042 \\
1999 Feb 24 && 2451234.647 & 1.550 & 0.019 && 2451234.631 & 2.088  & 0.024 \\
1999 Feb 25 && 2451235.680 & 1.562 & 0.011 && 2451235.673 & 2.055  & 0.011 \\
\bottomrule             
\end{tabular}
\end{table*}

In Table~\ref{table:observations-I} we report differential photometry of
HH~30 against the local standard in the instrumental $R$ and $I$
systems. In Table~\ref{table:observations-I}, $m$ is the relative magnitude of HH 30 (that is, the instrumental magnitude of HH~30 minus instrumental magnitude of the local standard) and $\sigma_m$ is the standard deviation in each relative magnitude estimated from photon statistics. The standard deviation of 
a single measurement about the mean is 0.242 in $R$ and 0.199 in $I$.
These are an order of magnitude larger than the variations seen in
Table~\ref{table:local-standard} between the local standard and four
field stars and an order of magnitude larger than the expected errors
from photon statistics. This suggests that the variability of HH~30 in
data set 1 is real.

\subsection{Data Set 2}

\begin{table*}
\centering\footnotesize
\setlength{\tabnotewidth}{\columnwidth}
\tablecols{13}
\caption{Data Set 2: Photometry of HH~30 from 1999 September to 2000 February}
\label{table:observations-II}
\begin{tabular}{lcccccccccccc}
\toprule
&&
\multicolumn{3}{c}{Instrumental $V$}&&
\multicolumn{3}{c}{Instrumental $R$}&&
\multicolumn{3}{c}{Instrumental $I$}\\
\cline{3-5}
\cline{7-9}
\cline{11-13}
\multicolumn{1}{c}{Date}&&
JD & $m$& $\sigma_m$&&
JD  & $m$& $\sigma_m$&&
JD  & $m$& $\sigma_m$\\
\midrule
1999 Sep 07 && 2451428.9921  & 14.207  & 0.027   && 2451428.9960  & 13.390  & 0.039   && 2451428.9985  & 13.506  & 0.040   \\
            && 2451429.0017  & 14.184  & 0.032   && 2451429.0055  & 13.414  & 0.039   && 2451429.0080  & 13.520  & 0.040   \\
1999 Sep 08 && 2451429.9946  & 14.107  & 0.040   && 2451429.9984  & 13.345  & 0.052   && 2451430.0009  & 13.493  & 0.053   \\
            && \nodata       & \nodata & \nodata && 2451430.0077  & 13.360  & 0.045   && 2451430.0103  & 13.487  & 0.051   \\
1999 Sep 11 && 2451433.0002  & 13.838  & 0.041   && 2451433.0041  & 13.236  & 0.040   && 2451433.0066  & 13.415  & 0.041   \\
            && 2451433.0107  & 13.877  & 0.026   && 2451433.0146  & 13.264  & 0.033   && 2451433.0171  & 13.469  & 0.042   \\
1999 Sep 13 && 2451435.0059  & 14.076  & 0.037   && 2451435.0098  & 13.376  & 0.049   && 2451435.0123  & 13.551  & 0.053   \\
1999 Sep 14 && 2451435.9991  & 14.647  & 0.054   && 2451436.0030  & 13.883  & 0.049   && 2451436.0055  & 14.057  & 0.060   \\
1999 Sep 15 && 2451436.9653  & 14.982  & 0.046   && 2451436.9692  & 14.042  & 0.049   && 2451436.9717  & 14.275  & 0.066   \\
1999 Oct 07 && 2451459.0033  & 15.242  & 0.086   && 2451459.0072  & 14.155  & 0.028   && 2451459.0097  & 14.497  & 0.027   \\
            && 2451459.0131  & 15.182  & 0.023   && 2451459.0170  & 14.205  & 0.026   && 2451459.0195  & 14.506  & 0.028   \\
1999 Oct 08 && \nodata       & \nodata & \nodata && 2451460.0157  & 14.111  & 0.017   && 2451460.0182  & 14.513  & 0.022   \\
1999 Oct 09 && 2451460.9976  & 15.030  & 0.030   && 2451461.0018  & 14.090  & 0.021   && 2451461.0044  & 14.446  & 0.016   \\
1999 Oct 12 && 2451463.9639  & 14.355  & 0.020   && 2451463.9678  & 13.623  & 0.027   && 2451463.9703  & 13.826  & 0.024   \\
            && 2451463.9734  & 14.340  & 0.014   && 2451463.9772  & 13.634  & 0.022   && 2451463.9797  & 13.814  & 0.024   \\
            && 2451463.9829  & 14.361  & 0.016   && 2451463.9875  & 13.641  & 0.022   && 2451463.9900  & 13.828  & 0.020   \\
1999 Oct 30 && 2451481.9541  & 14.641  & 0.028   && 2451481.9580  & 13.787  & 0.020   && 2451481.9605  & 13.976  & 0.018   \\
            && 2451481.9663  & 14.625  & 0.037   && 2451481.9702  & 13.741  & 0.015   && 2451481.9727  & 14.008  & 0.024   \\
            && 2451481.9753  & 14.656  & 0.046   && 2451481.9792  & 13.757  & 0.019   && 2451481.9817  & 14.001  & 0.017   \\
            && \nodata       & \nodata & \nodata && 2451481.9881  & 13.771  & 0.019   && \nodata       & \nodata & \nodata \\    
1999 Nov 05 && 2451487.9283  & 14.980  & 0.058   && 2451487.9322  & 13.996  & 0.067   && 2451487.9347  & 14.277  & 0.072   \\
1999 Dec 03 && 2451515.7883  & 14.676  & 0.020   && \nodata       & \nodata & \nodata && 2451515.7947  & 14.078  & 0.018   \\
1999 Dec 08 && 2451520.7739  & 14.814  & 0.029   && 2451520.7778  & 13.991  & 0.029   && 2451520.7803  & 14.266  & 0.036   \\
            && 2451520.7929  & 14.816  & 0.031   && 2451520.7967  & 14.043  & 0.036   && 2451520.7992  & 14.357  & 0.041   \\
2000 Jan 23 && \nodata       & \nodata & \nodata && 2451566.6145  & 13.345  & 0.041   && 2451566.6167  & 13.627  & 0.041   \\
2000 Jan 26 && 2451569.6053  & 14.014  & 0.041   && 2451569.6000  & 13.264  & 0.041   && 2451569.6020  & 13.471  & 0.041   \\
2000 Jan 30 && 2451573.5785  & 14.534  & 0.041   && 2451573.5729  & 13.660  & 0.041   && 2451573.5754  & 13.782  & 0.041   \\
2000 Feb 28 && 2451602.6122  & 14.899  & 0.020   && 2451602.6153  & 13.990  & 0.029   && 2451602.6179  & 14.320  & 0.024   \\
\bottomrule             
\end{tabular}
\end{table*}

Wood et al.\ (2000) report observations of HH~30 with Harris $VRI$
filters at the 1.2 meter telescope of the F. L. Whipple Observatory on
18 nights between 1999 September 7 and 2000 February 28. HH~30 was
observed more than once on 7 of these nights. These authors kindly made
available their reduced photometry, and we reproduce it in
Table~\ref{table:observations-II} for posterity. The magnitudes in
Table~\ref{table:observations-II} are instrumental magnitudes. The
standard deviation of a single measurement about the mean is
0.408 in $V$, 0.315 in $R$, and 0.378 in $I$.

\subsection{Data Set 3}

\begin{table*}
\centering\footnotesize
\setlength{\tabnotewidth}{\columnwidth}
\tablecols{13}
\caption{Data Set 3: Photometry of HH~30 from 2005 September to 2006 February}
\label{table:observations-III}
\begin{tabular}{lcccc}
\toprule
&&
\multicolumn{3}{c}{$I$}\\
\cline{3-5}
\multicolumn{1}{c}{Date}&&
JD  & $m$& $\sigma_m$\\
\midrule
2005 Sep 11&& 2453624.97& 16.86 &0.05\\
2005 Sep 13&& 2453626.95& 17.53 &0.05\\
2005 Sep 15&& 2453628.95& 17.25 &0.05\\
2005 Sep 16&& 2453629.93& 16.86 &0.05\\
2005 Sep 17&& 2453630.95& 16.85 &0.05\\
2005 Sep 18&& 2453631.94& 17.10 &0.05\\
2005 Sep 19&& 2453632.93& 17.16 &0.05\\
2005 Sep 25&& 2453638.94& 16.48 &0.05\\
2005 Sep 26&& 2453639.94& 16.73 &0.05\\
2005 Oct 22&& 2453665.87& 17.71 &0.05\\
2005 Oct 23&& 2453666.84& 17.40 &0.05\\
2005 Oct 24&& 2453667.85& 17.02 &0.05\\
2005 Oct 26&& 2453669.85& 17.01 &0.05\\
2005 Oct 27&& 2453670.85& 17.25 &0.05\\
2005 Oct 29&& 2453672.90& 17.29 &0.05\\
2005 Oct 31&& 2453674.84& 16.83 &0.05\\
2005 Nov 04&& 2453678.87& 17.00 &0.05\\
2005 Nov 05&& 2453679.87& 17.15 &0.05\\
2005 Nov 06&& 2453680.90& 16.79 &0.05\\
2005 Nov 13&& 2453687.89& 17.43 &0.05\\
2005 Nov 14&& 2453688.85& 17.14 &0.05\\
2005 Nov 15&& 2453689.84& 16.67 &0.05\\
2005 Nov 19&& 2453693.81& 16.99 &0.05\\
2005 Nov 20&& 2453694.81& 17.46 &0.05\\
2005 Dec 10&& 2453714.77& 17.14 &0.05\\
2006 Feb 01&& 2453767.74& 16.68 &0.05\\
2006 Feb 02&& 2453768.72& 17.07 &0.05\\
2006 Feb 11&& 2453777.71& 17.13 &0.05\\
2006 Feb 12&& 2453778.70& 16.99 &0.05\\
\bottomrule             
\end{tabular}
\end{table*}

We observed HH~30 again with the 84 centimeter telescope of the Observatorio
Astronómico Nacional on Sierra San Pedro Mártir on 29 nights between
2005 September 11 and 2006 February 12. We used the POLIMA imaging
polarimeter (Hiriart et al.\ 2005) with the SITe1 $1024 \times 1024$ CCD
binned $2\times2$ and the observatory's $I$ (4 mm RG9) filter. In this
paper we present photometric results from these observations; see
Durán-Rojas et al.\ (2008) for more details and for polarimetric
results.

The POLIMA instrument has a rotating Glan-Taylor prism that serves as a
polarizing filter. Each night we obtained exposures of HH~30 with the
prism orientated at 0{\arcdeg}, 45{\arcdeg}, 90{\arcdeg}, and
135{\arcdeg}. We typically obtained ten 120 second exposures per night at each
position during 2005 September and ten 300 second exposures per night at
each position after this. The image quality was typically $2''$ FWHM.
The nights we present here are those that were adequate for polarimetry,
which means that the transparency was stable over the whole night.
Therefore, it is likely that these nights were also photometric.

We reduced each image by subtracting an offset calculated from the
overscan, subtracting a residual bias image, and dividing by a twilight-sky flat field. We obtained instrumental magnitudes for HH~30 using
aperture photometry with an object aperture of diameter $8''$ and a
sky annulus with an inner diameter of $8''$ and an outer diameter of
$16''$. We averaged the instrumental magnitudes in the 0{\arcdeg} and
90{\arcdeg} images to produce a magnitude in the total intensity.

We obtained an indirect photometric calibration of each night. Each
night we observed the unpolarized standards Hiltner 960 and BD
$+59\arcdeg$ 389 (Schmidt, Elston, \& Lupie 1992). However, these
standards are not photometric standards. Therefore, on three nights we
observed standards from Landolt (1992) to determine the color terms for
the $I$ filter and the standard magnitudes of Hiltner 960 ($I = 9.07
\pm0.01$) and BD $+59\arcdeg$ 389 ($I=7.49\pm0.01$). We then calibrated
the photometry of HH~30 using a zero point determined for each night
from our observations of these standards, a color correction assuming
the color coefficient determined from our observations of Landolt
standards and a color of $V-I =1.82$ for HH~30 (Watson \& Stapelfeldt 2007), and
an atmospheric extinction correction using the mean extinction curve of
Schuster \& Parrao (2001). The uncertainties in our magnitudes are thus
dominated by systematic errors in this process, and we estimate them to
be roughly 0.05 magnitudes. The standard deviation of a single
measurement about the mean is 0.282 in $I$. This is much larger than the
estimated uncertainty in single measurement. Our photometry is shown in Table~\ref{table:observations-III}. 

\section{Distribution Analysis}

\begin{figure*}
\centering
\includegraphics[width=0.66\linewidth]{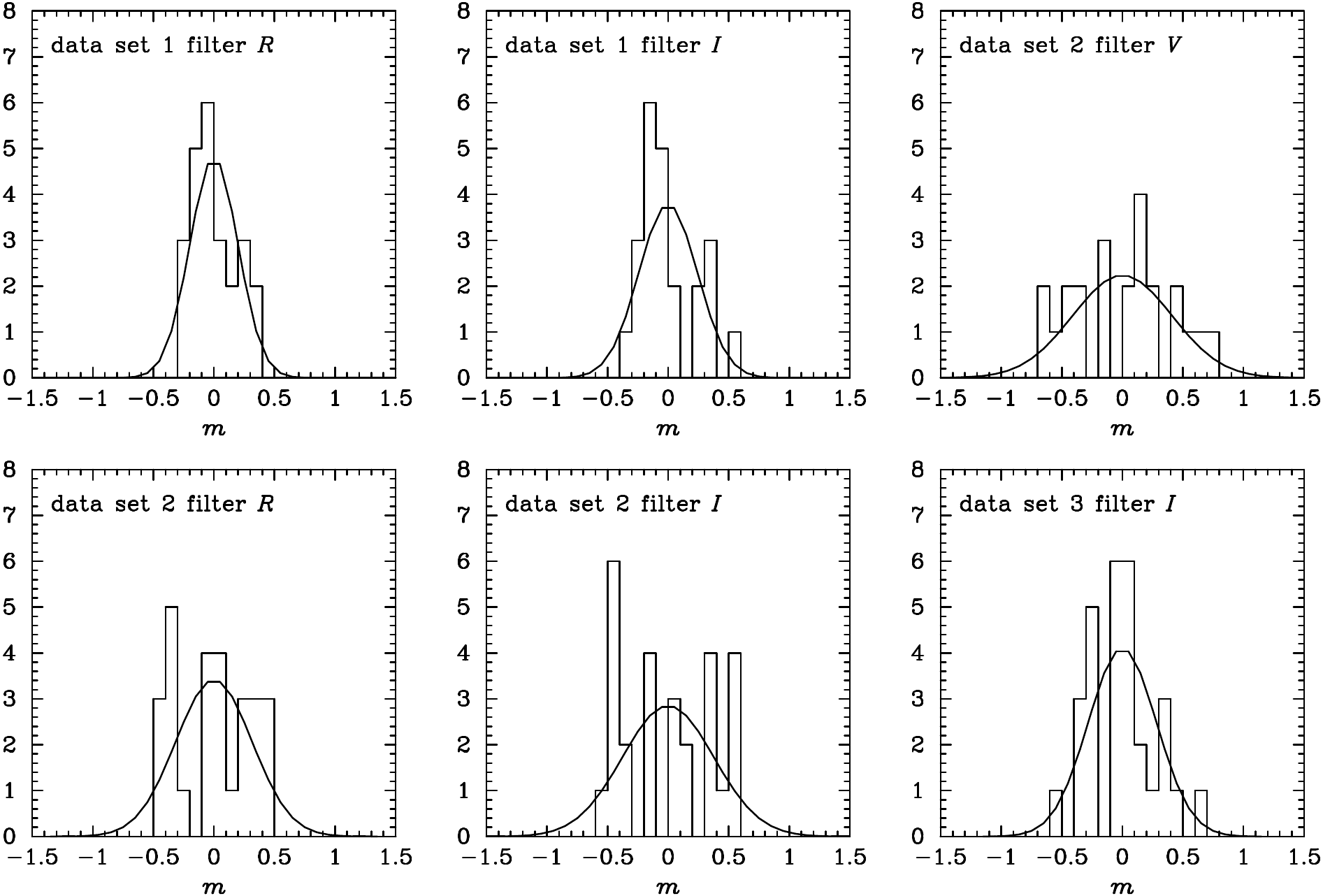}
\caption{Distributions of the data about their means. Each panel shows
  the histogram for the data and a Gaussian distribution with the same
  standard deviation. K-S tests show that the distributions are
  consistent with Gaussian distributions.}
\label{figure:distribution}
\end{figure*}

In \S\ref{section:period-analysis} we will use a null hypothesis that the data are independent and drawn from a Gaussian distribution. We will wish to use the rejection of this null hypothesis as evidence that the data are not independent. However, the data can fail this null hypothesis if they are not drawn from a Gaussian distribution. Therefore, we must first show that the data are indeed consistent with being drawn from a Gaussian distribution.

Figure~\ref{figure:distribution} shows the distributions of the data
about their means. Kolmogorov-Smirnov tests suggests that the null
hypothesis that the data sets are drawn from Gaussian distributions with
the same mean and standard deviation should be accepted with confidences
of 0.76 (data set 1 filter $R$), 0.52 (data set 1 filter $I$), 0.91
(data set 2 filter $V$), 0.48 (data set 2 filter $R$), 0.59 (data set
2 filter $I$), and 0.95 (data set 3 filter $I$). Thus, all of the
data sets are quite consistent with being drawn from Gaussian distributions.

\section{Period Analysis Methodology}
\label{section:period-analysis}

\subsection{The Lomb-Scargle Normalized Periodogram}

We have investigated the presence of a periodic signal in the data using
the Lomb-Scargle normalized periodogram (Lomb 1976; Scargle 1982; Press et al.\ 1992, \S
13.8). Periodic signals tend to create peaks in the periodogram.


The data sets are characterized by separations close to multiples of 1
day and as such contain little information below the corresponding
Nyquist period. Therefore, we searched for periods between 2 days and
half of longest separation present in each data set (which would allow
us to see two complete periods). We calculated the periodogram for 1000
periods per decade spaced evenly in the logarithm.

We characterized the significance of peaks in the periodogram against
the null hypothesis that the data points were independent and drawn from
a Gaussian distribution with mean $\bar m$ and variance $\sigma^2$. We
generated 10,000 trials under this null hypothesis and determined the
50\%, 90\%, 95\%, and 99\% confidence levels.

\subsection{Problems with Short-Term Correlations}

HH~30 shows short-term photometric correlations. For example, the largest
intra-night peak-to-valley variability in $I$ in data set 2 is 0.054
magnitudes (on the night of 1999 September 11), whereas the global
standard deviation is 0.38 magnitudes. Less dramatically, the standard
deviation in the difference of the $I$ magnitude between one night and
the previous night in data set 1 is 0.16 magnitudes whereas the standard
deviation of the $I$ magnitude of the same nights is 0.19 magnitudes.

\begin{figure*}
\centering
\includegraphics[width=\linewidth]{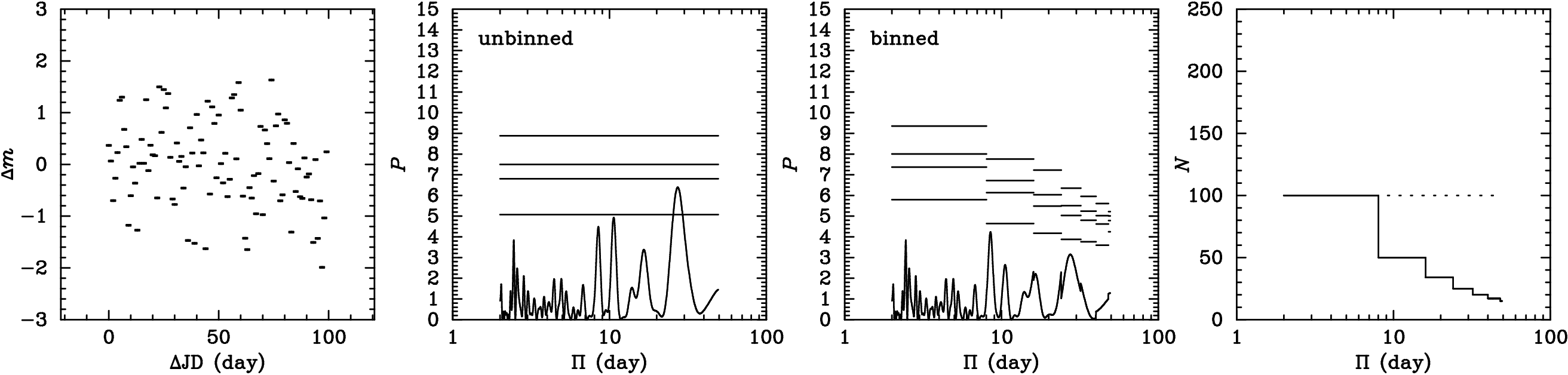}
\caption{Synthetic data and the resulting periodograms for a source that varies from night to night but
  is constant within a given night. The
  source is observed once per night. The first panel shows the data as a time series. The second panel shows the periodogram of the unbinned data and the 50\%, 90\%, 95\%, and 99\% confidence levels. The third panel shows the periodogram of the binned and the 50\%, 90\%, 95\%, and 99\% confidence levels. The last panel shows the number of effective data points in the periodogram of the binned data, with the dotted line referring to the unbinned data and the solid line referring to the binned data. }
\label{figure:false-1}
\end{figure*}

\begin{figure*}
\centering
\includegraphics[width=\linewidth]{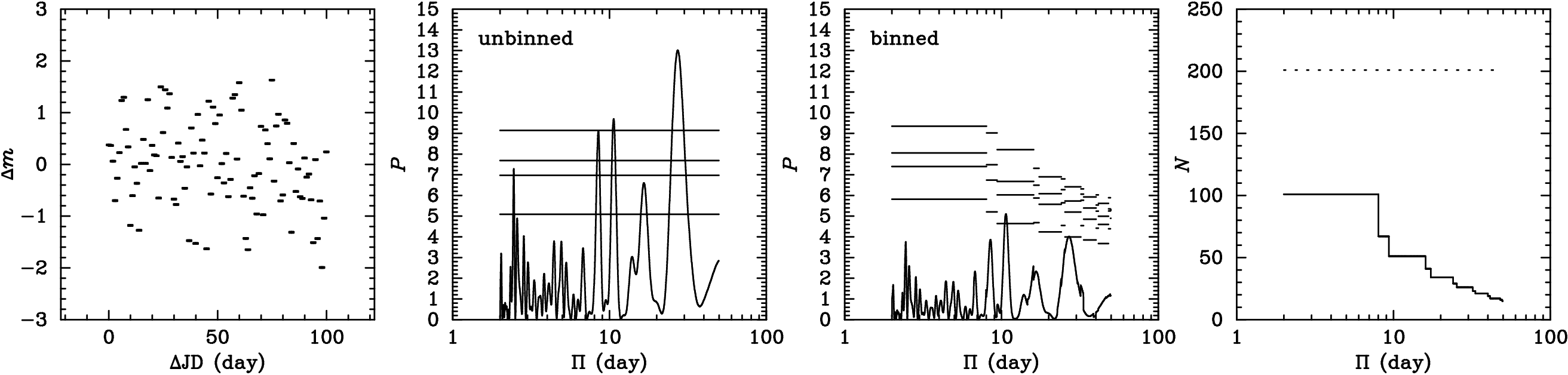}
\caption{Synthetic data and the resulting periodograms for a source that varies from night to night but
  is constant within a given night. The
  source is observed twice per night. The panels are as in
  Figure~\ref{figure:false-1}. Comparing this to
  Figure~\ref{figure:false-1}, it is clear that the peaks in the
  unbinned periodogram do not correspond to periodic signals and that binning reduces them
  appropriately. }
\label{figure:false-2}
\end{figure*}

Short-term correlations can cause problems for period searches using the
Lomb-Scargle normalized periodogram (Herbst \& Wittenmyer 1996). 
To see this, consider a hypothetical source
that varies in such a way that its magnitude over a single night is
constant but the magnitude for a given night is independent of the other
nights. Such as source exhibits perfect short-term correlation and
perfect long-term independence.

Consider observing this source once per night every night for 101
nights. Furthermore, consider that the observations are noiseless. The
first and second panels of Figure~\ref{figure:false-1} show an example
realization of this experiment and the corresponding periodogram
calculated from periods of 2 days, the Nyquist period, up to 50 days,
the longest period for which we could see two periods in the data. In
the plot of the periodogram the horizontal lines indicate the 50\%,
90\%, 95\%, and 99\% confidence levels. As expected, there are no
significant peaks in the periodogram; we accept the null hypothesis that
the data are independent.

Now consider observing the same source twice per night, with
observations separated by one hour. This generates two identical
magnitudes for each night. The first and second panels of
Figure~\ref{figure:false-2} show an example realization of this
experiment (with the same nightly magnitudes as
Figure~\ref{figure:false-1}) and the corresponding periodogram. The
structure in the periodogram is the same as in the
Figure~\ref{figure:false-1}, but the peaks are higher by roughly a
factor of two. This is expected from the expression for the periodogram
with duplicate data. However, the values corresponding to the different
confidence levels have hardly changed. Two of the peaks lie above the
99\% confidence level, and, on this basis, we strongly reject the null
hypothesis. This is not surprising; the null hypothesis is that the data
are independent, but half of the data are equal to the other half and so
clearly are not independent. However, we cannot interpret this rejection
as evidence for the presence of a period.

Thus, short-term correlations can generate peaks in the periodogram that
mimic those generated by periodic signals. Furthermore, periodic signals
are correlated over intervals that are short compared to the period.
Thus, a periodic signal that is finely sampled will have peaks in the
periodogram that arise both from short-term correlations and from the
periodic signal.

\subsection{Mitigating Short-Term Correlations}

We would like to distinguish peaks caused by short-term
correlations from peaks caused by periodic signals. The most rigorous
solution would probably be to use a null hypothesis that incorporated
the short-term correlations in the data. 

In a series of studies of stellar
variability in the Orion Nebula, Stassun et al.\ (1999) use a null hypothesis with two Gaussians, one for intra-night variability and one for inter-night variability, Rebull (2001) uses a null hypothesis with correlated Gaussian noise, and Herbst et al.\@ (2002)
essentially modify the null hypothesis from ``the data are independently
distributed'' to ``the data are similar to the same data with the
individual nights shuffled randomly''. These methods works work well 
provided one understands the timescale over
which correlations occur. The photometric variability of young stars
typically shows strong intra-night correlations but
only weak inter-night correlations, so shuffling whole nights is appropriate.
However, if the correlation were shorter or longer, one would need to
shuffle groups of data shorter than or longer than a single night.

In the case of HH~30 we are studying the photometric variability of a
young star, but one in which almost all of the light we see is scattered
by the circumstellar disk. It is not clear if the dominant variability in HH~30
is the same as in other young stars that are seen
directly. Therefore, we cannot assume that the correlations in HH~30 are
necessarily similar to those seen in other stars and cannot without
further investigation adopt the method of Herbst et al.\@ (2002).

Instead, we suggest a different means to mitigate short-term
correlations: we bin the data over intervals in which they are likely to
be correlated if a periodic signal is present. We suggest binning the data in bins equal to a given
fraction $f$ of the period being tested. 
We use adaptive bins; we start the first bin at the first data point and start subsequent bins at the first data point after the end of the previous bin. This binning has to
be carried out anew for each period being tested. (An alternative
approach would be to simply reject data within a certain interval of
non-rejected data.) Even with binning, some correlations may well remain
in the data. However, these correlations should be identical for all
periodic signals that have the same form but different periods. In this
sense, this procedure is uniformly biased rather than completely
unbiased. 

We need to select a suitable value for $f$; we have
chosen 1/8 (i.e., we bin data in intervals covering 1/8 of a period).
This is coarse enough to remove much of the correlations in a periodic
signal but not too coarse as to completely eliminate the signal, at least
for relatively smooth modulations. For data set 2, we will investigate
other values of $f$ in \S\ref{section: Period Analysis Results}.

The third panels of Figures~\ref{figure:false-1} and
\ref{figure:false-2} show periodograms calculated after binning the 
data into bins of $1/8$ of the period. The fourth panel in each figure
shows the number of data points without binning as a dotted line and
with binning as a solid line. In Figure~\ref{figure:false-2}, even
though there are 202 unbinned measurements (two per night for 101 nights),
the number of effective measurements is always 101 or less, as each night's observations are separated by only 1
hour and the
minimum binning interval is 6 hours (i.e., 1/8 of the minimum tested
period of 2 days). At the longest tested periods, the number of effective
measurements is about 16, which corresponds to 101 nights binned into
intervals of about 6 days (i.e., 1/8 of the maximum tested period of 50
days). By binning, we ensure that the effective number of points at a
given period is approximately the same in both
Figures~\ref{figure:false-1} and \ref{figure:false-2}.

In the unbinned test, the confidence level is assumed to be independent of the period. Unfortunately, in the binned test, the confidence level is now a strong function of the
period being tested. To calculate the confidence levels using the standard Monte Carlo method, 
we assume
that the probability of a false positive is uniformly distributed in the
logarithm of the period, which allows us to calculate the appropriate confidence for each interval in which the binning is constant. The results are shown in the third panels of
Figures~\ref{figure:false-1} and \ref{figure:false-2} as stepped
horizontal lines at the 50\%, 90\%, 95\%, and 99\% confidence levels.
The periodograms for the binned data show the peaks at the same periods
as for the unbinned data, but none of the peaks is especially significant;
the highest peak in the third panel of Figure~\ref{figure:false-1} has a
significance of less than 90\%. Thus, by binning the data we have successfully eliminated the peaks created by short-term correlations.

\begin{figure*}
\centering
\includegraphics[width=\linewidth]{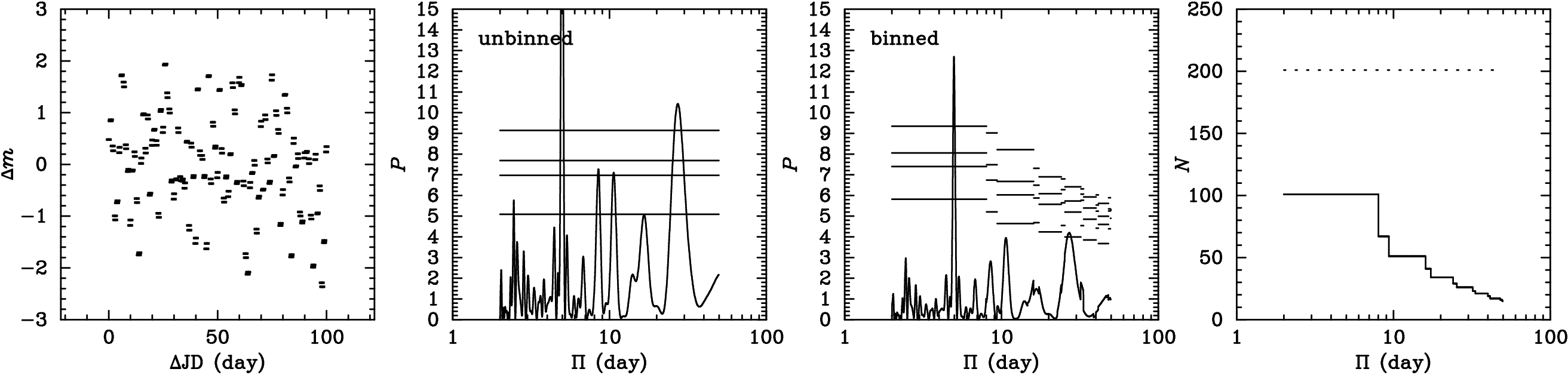}
\caption{Synthetic data  and
  the resulting periodograms for a source with two components: a noise component that varies from night to night but is constant within a night and a periodic component with a period of 5 days. The source is observed twice per night.
  The peak-to-valley amplitude of the periodic component is equal to the
  standard deviation of the noise. The panels are as in
  Figure~\ref{figure:false-1}.}
\label{figure:false-2-period-5}
\end{figure*}

\begin{figure*}
\centering
\includegraphics[width=\linewidth]{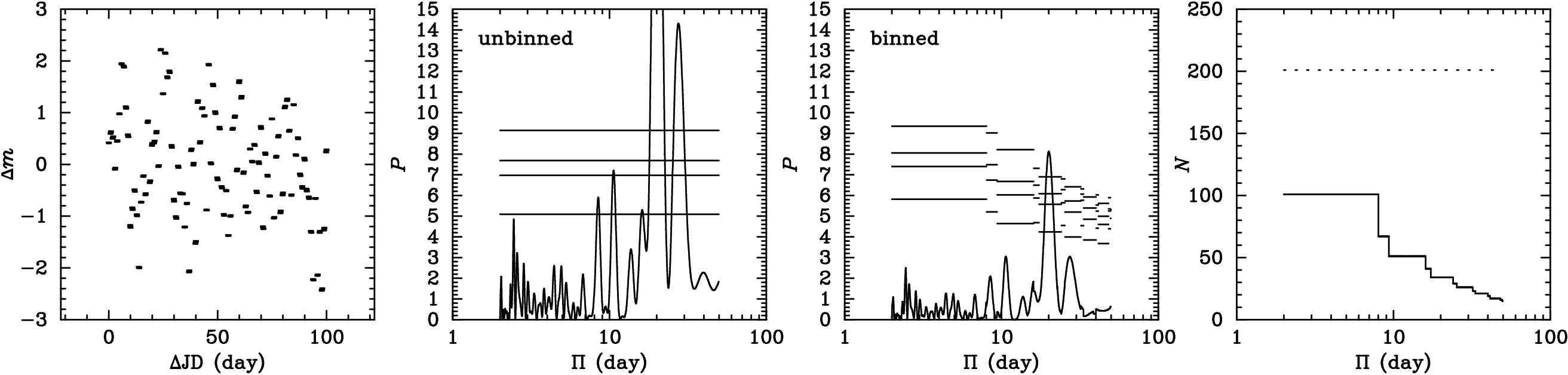}
\caption{Synthetic data  and
  the resulting periodograms for a source with two components: a noise component that varies from night to night but is constant within a night and a periodic component with a period of 20 days. The source is observed twice per night.
  The peak-to-valley amplitude of the periodic component is equal to 1.5 times
  the standard deviation of the noise. The panels are as in
  Figure~\ref{figure:false-1}.}
\label{figure:false-2-period-20}
\end{figure*}

Periodograms of binned data can still detect periodic signals.
Figures~\ref{figure:false-2-period-5} and
\ref{figure:false-2-period-20} show binned and unbinned periodograms for
data that are drawn from noisy periodic signals. To generate these, we added periodic component to the data used for Figure~\ref{figure:false-2}. In \ref{figure:false-2-period-5}, the period component had a period of 5 days and peak-to-valley amplitude equal to the standard deviation of the noise. In \ref{figure:false-2-period-20}, the periodic component had a period of 20 days and peak-to-valley amplitude equal to 1.5 times the standard deviation of the noise. The periodograms of the binned data correctly show the
period at 5 days in Figure~\ref{figure:false-2-period-5} and 20 days in
Figure~\ref{figure:false-2-period-20}.

\section{Period Analysis Results}
\label{section: Period Analysis Results}

\begin{figure*}
\centering
\includegraphics[width=\linewidth]{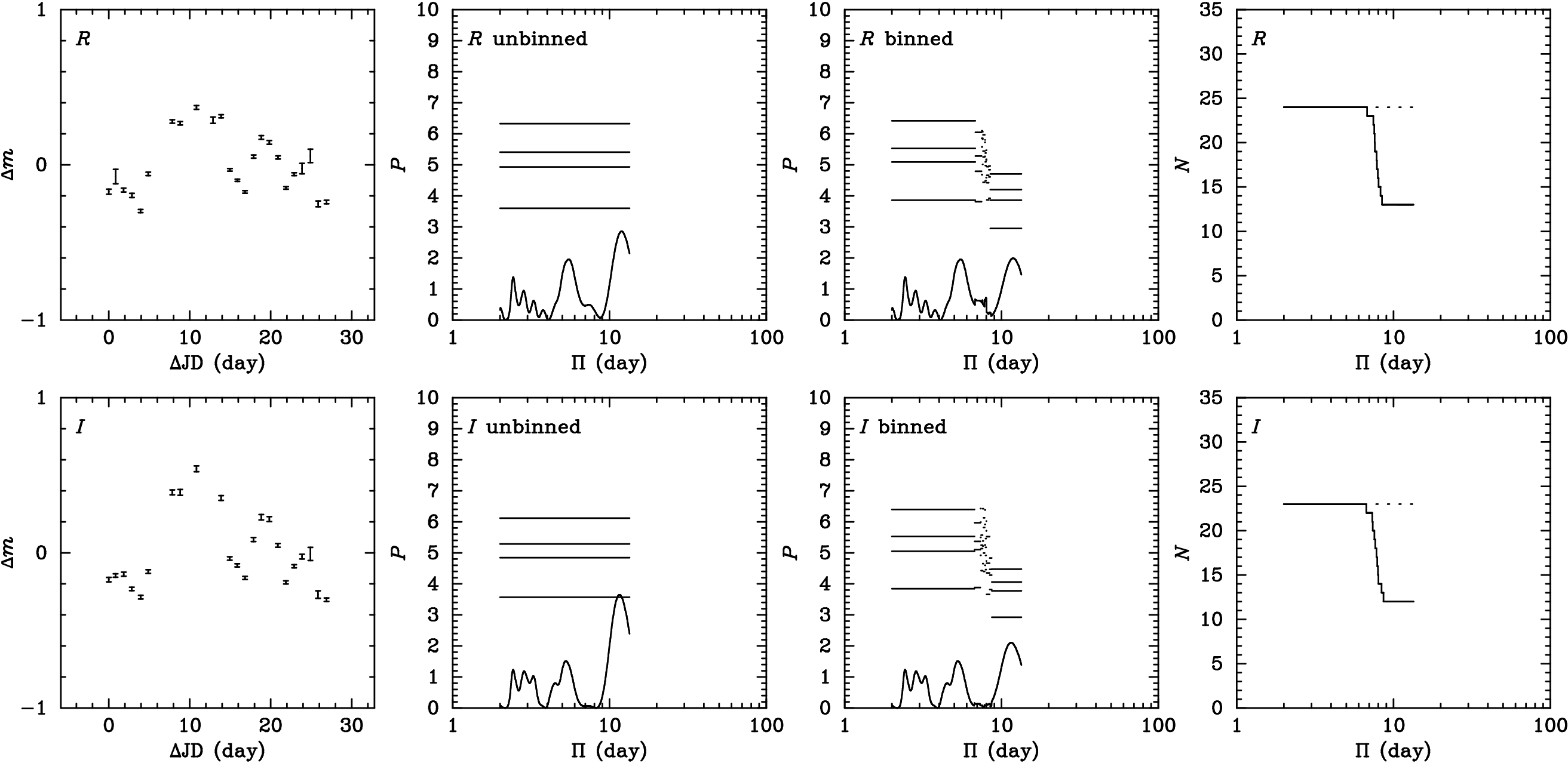}
\caption{Data set 1 and the resulting periodograms. Each row corresponds
  to a different photometric filter. In each row, the panels are as in
  Figure~\ref{figure:false-1}.}
\label{figure:data-set-I}
\end{figure*}

\begin{figure*}
\centering
\includegraphics[width=\linewidth]{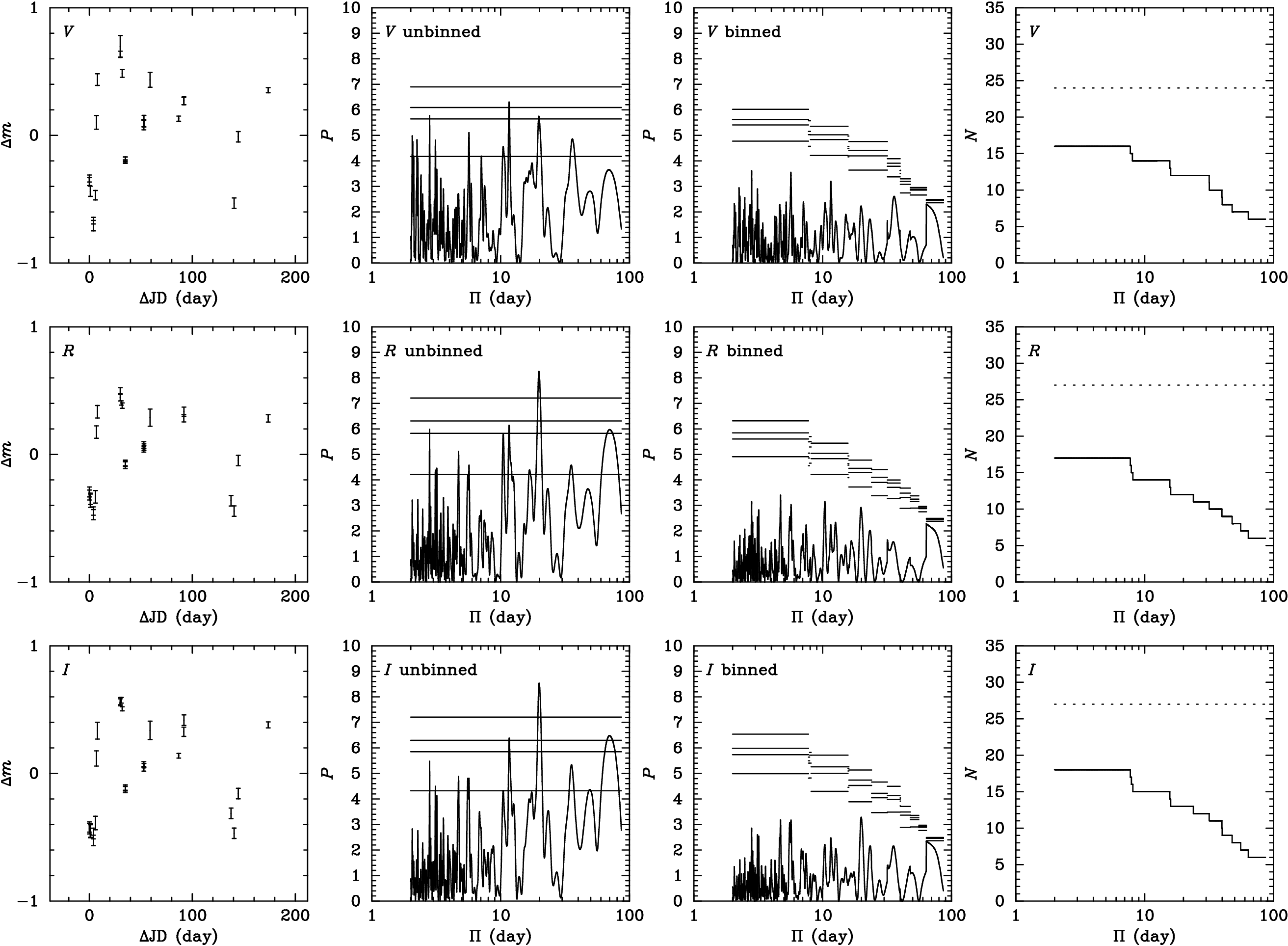}
\caption{Data set II and the resulting periodograms. Each row corresponds to
  a different photometric filter. In each row, the
  panels are as in Figure~\ref{figure:false-1}.}
\label{figure:data-set-II}
\end{figure*}

\begin{figure*}
\centering
\includegraphics[width=\linewidth]{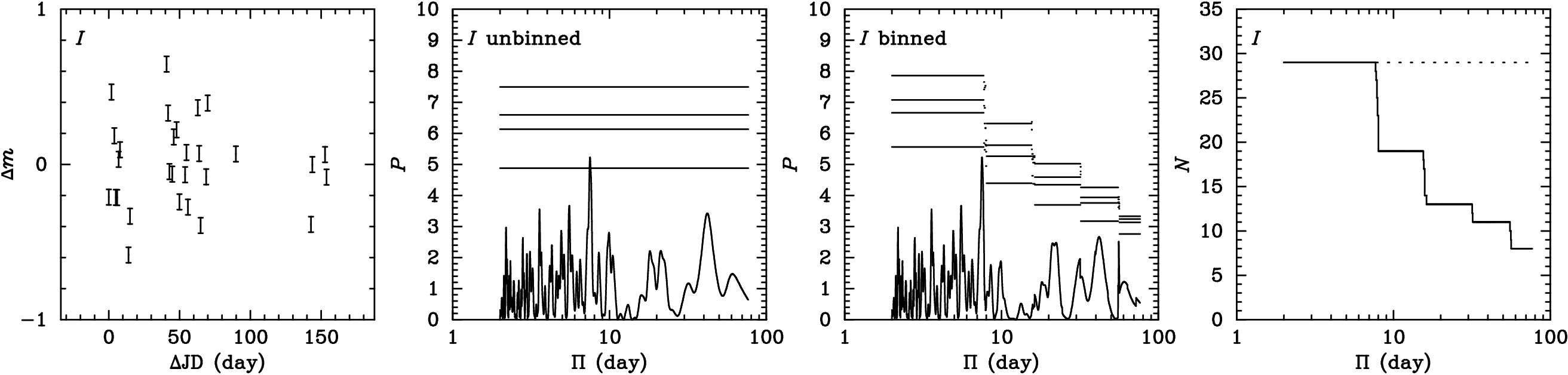}
\caption{Data set III and the resulting periodograms. The panels are as
  in Figure~\ref{figure:false-1}.}
\label{figure:data-set-III}
\end{figure*}

Figures~\ref{figure:data-set-I}, \ref{figure:data-set-II}, and
\ref{figure:data-set-III} show the data, periodograms, and number of
effective points for data sets 1, 2, and 3. The periodograms are
calculated at periods ranging from 2 days to 13 days (data set 1), 87 days
(data set 2), and 77 days (data set 3). As in
Figures~\ref{figure:false-1} and \ref{figure:false-2}, the first panel
in each row shows the data, the second panel the periodogram calculated
without binning, the third panel the periodogram with binning to $f =
1/8$ of the period, and the fourth panel the number of data points
without binning as a dotted line and with binning as a solid line. In
Figures \ref{figure:data-set-I} and \ref{figure:data-set-II}, each row
corresponds to observations in a different filter.

The unbinned periodograms for data sets 1 and 3 show no strongly
significant peaks. The highest peaks in $I$ have periods of 12.1 and 7.5
days and significances of only slightly more than 50\%. However, the
unbinned periodogram for data set 2 shows peaks at periods
of about 11.6, 19.9, and 69.9 days with significances in $I$ of more
than 95\%, 99\%, and 95\% respectively. These peaks also appear to be
present in $R$ at similar significances and in $V$ at reduced
significances. The first two periods were reported by Wood et
al.\ (2000).

However, the binned periodograms for data sets 1, 2, and 3 show no
significant peaks. The highest peak in $I$ in data set 2 is
still at 19.9 days but now with a significance of less than 50\%. It
appears that the strong peaks in the unbinned periodogram for data set
2 are entirely the result of short-term correlations in the data.

We mentioned above that the choice $f$, the bin size in units of the
period being examined, is open to some debate. We used $f = 1/8$ in the
figures and obtained no significant peaks in the periodogram. One might
ask if other values of $f$ might give different results. For example, in
Figure~\ref{figure:data-set-III}, one might wonder if a slightly larger
value of $f$ might increase the significance of the peak at about 7.5
days. In order to investigate the robustness of the lack of significant
peaks, we repeated the analysis with $f = 1/6$, 1/7, 1/8, 1/9, 1/10,
1/11, and 1/12, generating periodograms and confidence levels for each of these values. None of these periodograms showed a significant peak.

Recalculating the binned periodogram with $f=1/20$ yields a peak in $I$
in data set 2 at 19.9 days with a marginal significance of 90\%. However, in order
to accept this peak as indicative of a real periodic signal, we need to
accept that samples of a periodic signal separated by only $1/20$ of the
period are still effectively independent. This seems extremely unlikely.

We conclude that there is no significant evidence for a periodic photometric signal
in any of the data sets.




\section{Discussion}

\subsection{No Significant Periodic Photometric Variability}

Our analysis indicates that HH~30 shows photometric variability in $V$, $R$, and $I$ (as previously reported), but that periodograms show no significant evidence for a periodic photometric signal between periods of 2 and 87 days. This result is in disagreement with Wood et el.\ (2000); we suggest that correctly accounting for short-term correlations explains this difference. Of course, this result does not mean that there is no periodic photometric signal present. Rather, it simply means that any periodic signal must be sufficiently weak that it is hidden in the non-periodic noise.




\subsection{Origin of the Photometric Variability}

The large amplitude of the variability in $I$ (0.8, 1.1, 1.2, and 1.4 magnitudes in data sets 1, 2, and 3 and in the observations reported by Watson \& Stapelfeldt (2007) along with the lack of a detected period suggests that the photometric variability in HH~30 is related to Type II variability seen in other young stellar objects (Herbst et al.\ 1994). This is most common in classical T Tauri stars; Lamm et al.\ (2004) found that 61\% of the stars in NGC~2264 show this sort of irregular variability whereas only 31\% show significant periodic variability. Type II variability is thought to be caused by a variable accretion luminosity. This is consistent with the presence of strong collimated jets in HH~30 (Mundt et al.\ 1990; Burrows et al.\ 1996; Ray et al.\ 1996).

\subsection{Simultaneous HST Imaging}

Watson \& Stapelfeldt (2007) observed HH~30 with the WFPC2 camera of the {\it Hubble Space Telescope} on 1999 February 3, coincidentally during the period in which data set 1 was obtained. At this epoch, HH~30 showed a strong lateral
asymmetry in the upper nebula. The photometry of data set 1 shows that the magnitude of HH~30 was close to the minimum of its range at this time but rose to the maximum six days later. However, in the absence of evidence for a periodic photometric variability, we are not sure of the significance of these events.

\acknowledgements

We thank an anonymous referee for comments which helped improve the presentation of this work. We are extremely grateful to the staff of the OAN/SPM for their technical support and warm hospitality during several long observing runs. We thank David Hiriart,  Jorge Valdez, Fernando Quirós, Benjamín García, and Esteban Luna for their contributions to the design and construction of POLIMA. MCDR thanks CONACyT for a graduate student fellowship. KRS acknowledges support from HST GO grants 8289, 8771, and 9236 to the JPL/Caltech.

We used the IRAF software for some data reduction. IRAF is distributed by the National Optical Astronomy Observatories, which are operated by the Association of Universities for Research in Astronomy, Inc., under cooperative agreement with the National Science Foundation.

We also used data products from the Two Micron All Sky Survey, which is a joint project of the University of Massachusetts and the Infrared Processing and Analysis Center at the California Institute of Technology, funded by the National Aeronautics and Space Administration and the National Science Foundation.

\end{document}